\documentclass[10pt,conference]{IEEEtran}
\IEEEoverridecommandlockouts

\usepackage{cite}
\usepackage{amsmath,amssymb,amsfonts}
\usepackage{algorithmic}
\usepackage{graphicx}
\usepackage{textcomp}
\usepackage{xcolor}
\usepackage{booktabs}
\usepackage{hyperref}
\usepackage{siunitx}

\def\BibTeX{{\rm B\kern-.05em{\sc i\kern-.025em b}\kern-.08em
    T\kern-.1667em\lower.7ex\hbox{E}\kern-.125emX}}
\begin{document}

\title{MLScent: A tool for Anti-pattern detection in ML projects}

\author{\IEEEauthorblockN{Karthik Shivashankar}
\IEEEauthorblockA{\textit{Department of Informatics},
\textit{University of Oslo, Norway}\\
karths@ifi.uio.no}
\and
\IEEEauthorblockN{Antonio Martini}
\IEEEauthorblockA{\textit{Department of Informatics},
\textit{University of Oslo, Norway}\\
antonima@ifi.uio.no}}

\maketitle

\begin{abstract}
Machine learning (ML) codebases face unprecedented challenges in maintaining code quality and sustainability as their complexity grows exponentially. While traditional code smell detection tools exist, they fail to address ML-specific issues that can significantly impact model performance, reproducibility, and maintainability.

This paper introduces MLScent, a novel static analysis tool that leverages sophisticated Abstract Syntax Tree (AST) analysis to detect anti-patterns and code smells specific to ML projects.

MLScent implements 76 distinct detectors across major ML frameworks including TensorFlow (13 detectors), PyTorch (12 detectors), Scikit-learn (9 detectors), and Hugging Face (10 detectors), along with data science libraries like Pandas and NumPy (8 detectors each). The tool's architecture also  integrates general ML smell detection (16 detectors), and specialized analysis for data preprocessing and model training workflows. 

Our evaluation demonstrates MLScent's effectiveness through both quantitative classification metrics  and qualitative assessment via user studies feedback with ML practitioners. Results show high accuracy in identifying framework-specific anti-patterns, data handling issues, and general ML code smells across real-world projects.

\end{abstract}

\section{Introduction}

The software development landscape has undergone a dramatic transformation with the integration of Machine Learning (ML). Recent statistics from Gartner highlight this shift, revealing a striking 270\% increase in ML adoption within enterprise software projects over the last four years \cite{gartner2020}. This growth spans multiple sectors, with healthcare leading the charge \cite{marketsandmarkets2021}.

This rapid adoption, however, brings its own set of complexities. Traditional software development practices have had to evolve significantly to accommodate ML's unique requirements, including the need for extensive datasets, sophisticated algorithms, and iterative development cycles \cite{amershi2019}. These fundamental differences have catalyzed a complete reimagining of software development methodologies, from initial design through testing and maintenance \cite{oliveira2023challenges, di2024code} which is also highlighted by Tang et al. \cite{tang2021} in their empirical study of ML systems refactoring and technical debt.

ML projects introduce distinct code quality challenges that set them apart from conventional software development. The complexity stems from their inherent characteristics: intricate mathematical operations, extensive data preprocessing requirements, and sophisticated model architectures that challenge traditional code maintenance approaches \cite{sculley2015}. The iterative nature of ML development, where models undergo continuous refinement based on performance metrics, often results in code that becomes increasingly difficult to maintain and understand.

The consequences of subpar code quality in ML projects can be particularly severe. Beyond the usual software maintenance issues, poor code quality can compromise model performance, hinder result reproducibility, and create significant obstacles during production deployment \cite{breck2017}. O'Brien et al. \cite{OBrien2022} further elaborate on these challenges through their comprehensive study of self-admitted technical debt in ML software, identifying 23 distinct patterns of technical debt specific to ML systems. These challenges can severely impact team collaboration and make model updates problematic as new data becomes available \cite{gao2022challenges}.

ML code smells represent unique indicators of potential system design or implementation issues that traditional metrics might miss \cite{liu2019}. Zhang et al. \cite{zhang2022code} have extensively documented these ML-specific smells, which often manifest in ways distinct from conventional software development – such as the inappropriate use of default hyperparameters or improper data splitting techniques for training and testing.

The spectrum of ML-specific code smells extends to various critical areas, as documented by  Oort et al. \cite{oort2021} in their study of code smell prevalence in ML projects. These include inadequate handling of data imbalances, lack of proper feature scaling, and the selection of inappropriate evaluation metrics \cite{nair2020}. Sculley et al. \cite{sculley2015} further emphasize how these issues can lead to hidden technical debt in ML systems.

While traditional code analysis tools excel in general software development, they often prove inadequate for ML codebases due to their lack of domain-specific understanding \cite{zhang2020}. This limitation is particularly evident in the context of ML-specific issues, as demonstrated by Cruz et al. \cite{cruz2020} in their empirical study of bad smell detection techniques. For instance, these tools might overlook critical ML-specific issues, such as the inappropriate use of classification accuracy metrics on imbalanced datasets \cite{di2024code}.

This limitation in existing tools underscores the pressing need for ML-specific code quality analysis solutions. Tsoukalas et al. \cite{Tsoukalas2022} emphasize the importance of specialized tools that must comprehend ML-specific constructs, workflows, and best practices \cite{idowu2021, zhang2023machine}. Alahdab and Çalıklı \cite{alahdab2019empirical} further support this through their empirical analysis of hidden technical debt patterns in ML software.

MLScent is  tailored for ML development, designed to detect and address ML-specific code smells and anti-patterns. Unlike general-purpose code quality tools, MLScent provides context-aware suggestions and detection of ML specific smells and anti-patterns may cater to ML workflows, helping practitioners improve code quality, maintainability, and development efficiency.

\section{Background and Related Work}

Code smells, initially conceptualized by Kent Beck and popularized by Martin Fowler \cite{fowler1999refactoring}, have evolved beyond traditional software development concerns to encompass machine learning-specific challenges. 

While traditional code smells focus on issues like duplicated code, oversized classes, and inappropriate coupling \cite{yamashita2013exploring, das2019}, ML systems present unique quality challenges that existing detection tools like SonarQube \cite{sonarqube} and PMD \cite{pmd} fail to adequately address \cite{barbez2019, oort2021}. These traditional tools, while effective for conventional software development \cite{fontana2012automatic}, lack the capability to comprehend ML-specific contexts and the nuances of ML frameworks and libraries.

ML-specific code smells, as documented by Sculley et al. \cite{sculley2015hidden, scully2014}, include data leakage issues, improper feature scaling, and inadequate cross-validation practices. These issues, identified through empirical studies by Tang et al. \cite{tang2021} and O'Brien et al. \cite{OBrien2022}, can significantly impact model performance and reliability. Additional concerns include improper train-test split implementations, the use of hardcoded hyperparameters without proper tuning, and poor handling of imbalanced datasets, as highlighted by Takeuchi et al. \cite{takeuchi2023practice} and Washizaki et al. \cite{washizaki2020machine}. 

The limitations of traditional code smell detection tools in understanding ML-specific contexts \cite{cruz2020} emphasize the need for specialized quality assurance approaches in ML development, particularly as ML systems become more complex and widespread. This growing challenge, further elaborated by Tsoukalas et al. \cite{Tsoukalas2022} and Alahdab and Çalıklı \cite{alahdab2019empirical}, underscores the importance of developing new tools and methodologies specifically designed to address the unique quality concerns in ML software development.

\subsection{Current State of ML Code Quality Tools}

The evolution of code quality tools has primarily focused on traditional software development, leaving a significant gap in tools specifically designed for machine learning applications. While several tools offer partial solutions for ML code quality assessment, they fall short of providing comprehensive coverage for the unique challenges posed by ML development.

In the current ML quality landscape, tools like MLlint \cite{vanOort2022} provide project-level quality analysis focusing on code, data, and configuration aspects. While valuable for general project assessment, it lacks the deep framework-specific analysis that MLScent offers. Data Linter \cite{dataLinter} and Datalab \cite{datalab} concentrate on dataset quality and auditing, addressing issues like noisy labels and outliers, but do not cover code-level anti-patterns specific to ML implementations.

More specialized tools like Deepchecks \cite{deepchecks} focus on model validation and performance checks, while Great Expectations \cite{greatExpectations} emphasizes data validation and profiling. MLflow \cite{Chen2020} excels in experiment tracking and model management but does not address code quality issues. Similarly, DVC \cite{dvc} specializes in version control for datasets and models, while Pandas Profiling \cite{pandasProfiling} focuses solely on data analysis visualization.

MLScent distinguishes itself from these existing solutions in several key ways:

Framework Coverage: Unlike tools that focus on specific aspects (e.g., MLflow for experiment tracking or DVC for version control), MLScent provides comprehensive coverage across multiple ML frameworks, including PyTorch, TensorFlow, Scikit-learn, and Hugging Face libraries like Transformers.

Integrated Analysis: While tools like mllint \cite{vanOort2022} offer general project quality assessment, MLScent combines framework-specific analysis with general ML anti-pattern detection, providing a more complete quality assessment solution.

Code-Level Focus: Unlike data-centric tools such as Data Linter \cite{dataLinter} or Deepchecks \cite{deepchecks}, MLScent specifically targets code-level issues, helping developers identify and fix implementation problems that could affect model performance and maintainability.

Extensibility: MLScent's architecture allows for easy integration of new framework-specific detectors, making it more adaptable to the evolving ML ecosystem compared to more rigid tools like Pylint or framework-specific analyzers.

These existing solutions, while valuable for their specific purposes, demonstrate significant limitations in their scope and applicability. They typically address isolated aspects of ML development or focus on specific frameworks, failing to provide the comprehensive analysis required for modern ML projects. The ideal ML code quality tool would need to comprehend multiple ML frameworks, detect framework-specific code smells, provide contextual improvement suggestions, and seamlessly integrate into existing ML development workflows - requirements that MLScent specifically addresses.

The development of MLScent is motivated by several critical gaps in the current landscape of ML code quality tools and research. A fundamental limitation identified by Kaur et al. \cite{kaur2021review} is the scarcity of ML-specific code smell detection tools, despite the abundance of such tools for traditional software development. While existing tools like DVC \cite{dvc}, MLflow \cite{Chen2020}, and Deepchecks \cite{deepchecks} excel in specific aspects like version control, experiment tracking, and data validation, they fail to provide comprehensive coverage of ML code quality issues.

The challenges are further compounded by the fragmented coverage of ML frameworks, as highlighted by Chen et al. \cite{Chen2020}, and the shortage of high-quality, industry-relevant datasets for ML code smell detection \cite{vanOort2022}. Additionally, there exists a growing disconnect between established software engineering practices and the unique requirements of ML development \cite{Ratner2017}.

Current tools like Great Expectations \cite{greatExpectations} and Pandas Profiling \cite{pandasProfiling}, while useful for specific data quality aspects, do not address the broader spectrum of ML-specific code smells. This gap is particularly noteworthy given the limited research into ML-specific code smells, with most existing studies focusing on conventional software development patterns \cite{madeyski2020mlcq}. MLScent aims to bridge these gaps by providing a comprehensive solution that aligns traditional software engineering principles with the specific demands of ML development workflows.

The comparative analysis presented in Table \ref{tab:ml_tools_expanded} highlights these distinctions, showing how MLScent provides unique coverage across general ML, data science, and framework-specific concerns. This comprehensive approach positions MLScent as a more complete solution for ML code quality assessment, addressing the limitations of existing tools while providing the specialized analysis needed for modern ML development.

\begin{table*}[t]
\centering
\caption{Comparison of ML tools for detecting code smells, anti-patterns, and quality issues. {\footnotesize Type: OS=Open Source, Res.=Research; Gen. ML=General Machine Learning support; Data Sci.=Data Science support; Frame. Spec.=Framework Specific features.}}
\label{tab:ml_tools_expanded}
\small
\setlength{\tabcolsep}{4pt}
\begin{tabular}{@{}llllccc@{}}
\toprule
\textbf{Tool} & \textbf{Type} & \textbf{Focus} & \textbf{Key Features} & \textbf{Gen. ML} & \textbf{Data Sci.} & \textbf{Frame. Spec.} \\
\midrule
mllint \cite{vanOort2022} & OS & Proj. quality & Code/data/config analysis; recommendations & \checkmark & \checkmark & \\
Data Linter \cite{dataLinter} & Res. & Dataset sanity & Issue ID, feature transforms & & \checkmark & \\
Datalab \cite{datalab} & OS & Data auditing & Noisy labels, outliers, duplicates; diagnostics & & \checkmark & \\
Deepchecks \cite{deepchecks} & OS & Model & Data valid., perf. checks, workflow integ. & \checkmark & \checkmark & \\
Great Expect. \cite{greatExpectations} & OS & Data valid. & Valid., profiling, custom expectations & & \checkmark & \\
MLflow \cite{Chen2020} & OS & Exp. tracking & Track exps., manage models, deployment & \checkmark & & \\
DVC \cite{dvc} & OS & Version control & Dataset/model versioning, reproducibility & \checkmark & & \\
Pandas Prof. \cite{pandasProfiling} & OS & Data analysis & Dataset reports, correlations, visualizations & & \checkmark & \\
MLScent& OS&  Anti-pattern and smells & Anti-pattern, smells ML checker & \checkmark& \checkmark& \checkmark\\
\bottomrule
\end{tabular}

\end{table*}

\section{MLScent: System Architecture and Design}

\subsection{Overall System Architecture}
\begin{figure}
    \centering
    \includegraphics[width=1\linewidth]{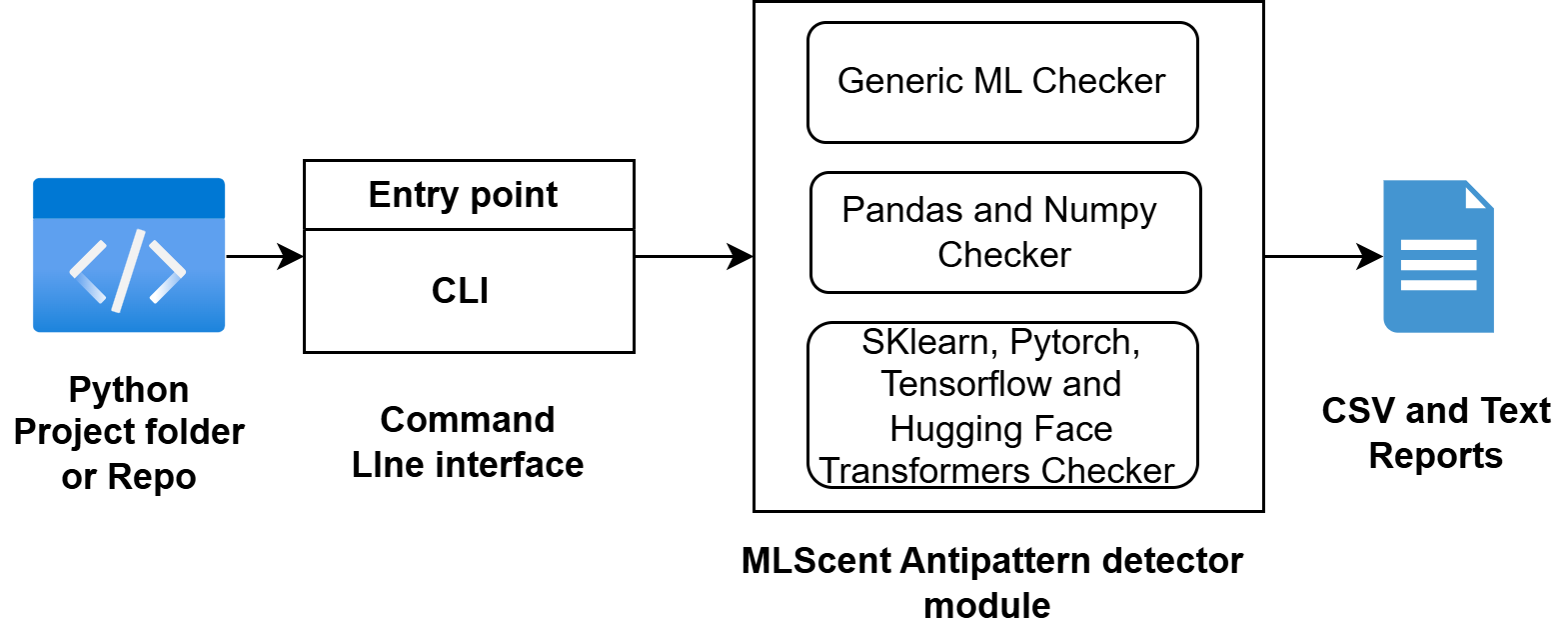}
    \caption{MLScent: Core Components}
    \label{fig:mlscent_architecture}
\end{figure}

MLScent employs a modular and extensible architecture designed specifically for detecting code smells in machine learning projects. As depicted in Figure \ref{fig:mlscent_architecture}, the system integrates several key components that work in harmony to deliver comprehensive code quality analysis across various ML frameworks.

The architecture consists of a Code Parser utilizing the Abstract Syntax Tree (AST) generation, Framework-Specific and Hugging Face Smell Detectors for specialized analysis, a General ML Smell Detector for framework-agnostic issues, and a Report Generator that produces both human-readable and machine-parsable outputs. 

The detectors in MLScent were built from scratch based on comprehensive analysis of ML anti-patterns and code smells documented in prior research. The smell definitions were derived from multiple sources, including Sculley et al.'s work on technical debt in ML systems \cite{sculley2015, OBrien2022}, Zhang et al.'s \cite{zhang2022code} catalog of ML-specific code smells , and van Oort et al.'s \cite{oort2021} study on ML code smell prevalence.

\subsection{Framework-Specific Smell Detector}

The Framework-Specific Smell Detector module in our library serves as a main component, focusing on identifying framework-specific code smells across popular ML libraries. Through AST analysis and pattern matching, this detector examines code for framework-specific anti-patterns and suboptimal structures. It implements specialized detection methods for frameworks like Pandas (checking for chain indexing issues), NumPy (identifying inefficient array operations), Scikit-learn (detecting cross-validation problems), and deep learning frameworks such as TensorFlow and PyTorch (analyzing model training practices). The detector's modular design facilitates easy extension to accommodate new frameworks and evolving best practices.

\subsection{Hugging Face Smell Detector}

The Hugging Face Smell Detector module in our library addresses the unique challenges in Natural Language Processing (NLP) projects using the Hugging Face library. This specialized component focuses on critical aspects such as model versioning, tokenizer caching, pipeline component usage, and training argument configuration. Its implementation is particularly valuable given the growing prominence of Hugging Face in NLP development, helping developers maintain high-quality, efficient, and reproducible NLP models.

\subsection{General ML Smell Detector}

The ML Smell Detector module identifies framework-agnostic code smells common across ML development. It examines crucial aspects including data handling and preprocessing, model training and evaluation practices, hyperparameter management, and overall code organization. This detector ensures comprehensive coverage even for projects using custom or less common ML frameworks, providing valuable insights regardless of the specific tools employed.

\subsection{Integration and Extensibility}
MLScent's architecture emphasizes seamless integration and extensibility through its main orchestration class. This class coordinates file parsing, detector invocation, result aggregation, and report generation. The system's modular design allows for straightforward extension through new detector implementation, refinement of existing detectors, and enhancement of reporting capabilities. This flexibility ensures MLScent can adapt to the evolving landscape of ML development while maintaining its effectiveness in promoting code quality.The tool uses AST to identify which ML framework is being used, and then automatically applies the appropriate framework-specific detectors, rather than running all detectors unnecessarily in the ML codebase or projects.

\subsection{AST-Based Analysis Techniques}
MLScent employs sophisticated Abstract Syntax Tree (AST) analysis techniques to detect code smells in machine learning projects. This approach provides a hierarchical representation of code structure, enabling efficient and accurate analysis of code patterns across various frameworks and libraries.

The AST-based analysis offers several key advantages in code smell detection. Its language-agnostic nature ensures consistent analysis across different Python versions, while contextual understanding capabilities enable accurate interpretation of code constructs. The efficient pattern matching through AST traversal allows for rapid identification of complex code patterns. 

However, the approach also faces certain limitations, including static analysis constraints and computational overhead with large codebases.

To address these challenges, MLScent combines AST analysis with heuristic-based detection and framework-specific rules for comprehensive code smell detection.

\subsection{Use of Astroid Library}

MLScent's implementation leverages the Astroid library as its primary tool for Python code parsing and analysis. This powerful library provides sophisticated AST functionality for deep code inspection and analysis.

The integration focuses on three key features: AST generation for code structure analysis, rich node types for precise pattern identification, and an inference engine for understanding variable types and function return values.

For instance lets consider Pandas examples for Anti-pattern checker in the Astroid library:

\paragraph{Pandas: Detecting Chain Indexing}

Chain indexing in Pandas can lead to unexpected behavior and performance issues. Here's how MLScent detects this smell:

\begin{scriptsize}
\begin{verbatim}
def detect_chain_indexing(node):
    if isinstance(node, astroid.Subscript):
        if isinstance(node.value, astroid.Subscript):
            return True
    return False

# Usage in MLScent
for node in astroid.parse(source_code).body:
    if detect_chain_indexing(node):
        report_smell("Chain indexing detected", node)
\end{verbatim}
\end{scriptsize}

This function checks for nested \texttt{Subscript} nodes, which indicate potential chain indexing like df['column']['row'].

These examples demonstrate how MLScent utilizes Astroid's AST parsing capabilities to perform sophisticated code analysis across various ML frameworks. By traversing the AST and checking for specific patterns, MLScent can identify potential code smells that are particularly relevant to ML development practices.

\begin{table}[htbp]
\centering
\caption{Count of  Anti-Pattern Detectors modules in MLScent Tool}
\begin{tabular}{lc}
\hline
Framework/Category & Number of Detectors \\
\hline
TensorFlow & 13 \\
General ML & 16 \\
PyTorch & 12 \\
HuggingFace & 10 \\
Scikit-learn & 9 \\
Pandas & 8 \\
NumPy & 8 \\
\hline
\textbf{Total} & \textbf{76} \\
\hline
\end{tabular}
\label{tab:mlscent_detectors}
\end{table}

MLScent is designed to detect anti-patterns and potential issues in machine learning code. The tool implements a total of 76 distinct detectors across major ML frameworks and categories as shown in Table \ref{tab:mlscent_detectors} and we also plan to extend this list with more detectors. The distribution of these detectors reflects the tool's extensive coverage of the ML ecosystem. MLScent particularly emphasizes General ML practices with 16 detectors, addressing common pitfalls in machine learning development regardless of the framework used. The tool provides robust coverage for modern deep learning frameworks, with 13 detectors for TensorFlow and 12 for PyTorch implementations. 

The inclusion of 10 HuggingFace-specific detectors demonstrates MLScent's attention to transformer-based models and natural language processing applications. 

For traditional machine learning workflows, MLScent implements 9 detectors for Scikit-learn. Data manipulation and processing are covered through 8 detectors each for Pandas and NumPy, ensuring comprehensive code quality analysis across the entire ML development pipeline. This broad distribution of detectors enables MLScent to provide thorough static analysis across different aspects of machine learning development, from data preprocessing to model deployment.

\subsection{Command-Line Interface}

MLScent implements a user-friendly command-line interface (CLI) leveraging Python's \texttt{argparse} library. This interface provides developers with a powerful and flexible tool for analyzing machine learning projects while maintaining simplicity in operation.

The CLI's design incorporates several essential features to enhance user experience and analysis capabilities. Users can perform granular analysis by specifying individual files or conduct comprehensive scans of entire project directories. The interface supports multiple output formats, allowing users to generate reports in both human-readable TXT format and machine-parsable CSV format for further processing. Additionally, the CLI offers fine-grained control over framework-specific analysis, enabling users to focus on particular ML frameworks relevant to their projects.

\subsection{Replication Package}
To ensure the reproducibility of our results and to facilitate further research in this area, we have prepared a comprehensive replication package for \texttt{MLScent}. This package includes all the necessary components to recreate our experiments and verify our findings.

The replication package is available through multiple channels:

\textbf{Documentation:} Detailed documentation of \texttt{MLScent}, including installation instructions, usage guidelines, and API references, is available at\footnote{\url{https://ml-scent.readthedocs.io/en/latest/}}.

\textbf{Source Code:} The complete source code of \texttt{MLScent}, is accessible \footnote{\url{https://github.com/KarthikShivasankar/ml_smells_detector}}.

\textbf{Data and Results:} To ensure long-term availability and citability, we have archived everything, including the analyzed ML projects, on Zenodo\footnote{\url{https://doi.org/10.5281/zenodo.14097743}}.
By providing this comprehensive replication package, we aim to contribute to the transparency and reproducibility of software engineering research in the machine learning domain.

\section{Evaluation and Results}

This section presents a comprehensive evaluation of MLScent, detailing the methodology used, the datasets analyzed, and both quantitative and qualitative results of our analysis.

\subsection{Evaluation Methodology}

To assess the effectiveness of MLScent, we employed a multi-faceted evaluation approach:

\begin{enumerate}
    \item  Classifcation: To establish a reliable benchmark for evaluating our ML-specific code smell detection tool (MLScent), we collaborated with the developers primarily responsible for the projects listed in Table \ref{tab:survey_responses}.  We asked each developer to annotate a randomly selected subset of code from their own projects, thus creating a "ground truth" dataset.

Before they began annotating, we provided the developers with detailed information about the types of ML-specific code smells we were targeting. This ensured they were well-prepared to identify potential issues within their code. By comparing the code smells detected by MLScent with the annotations provided by the developers, we were able to calculate key performance metrics such as recall, as well as F1 and F2 scores, which provide a more balanced measure of accuracy.
       
    \item User Study: We conducted a small-scale user study with same ML practitioners in Table \ref{tab:survey_responses}  to gather feedback on the tool's usability and the relevance of detected smells.
\end{enumerate}

For each detected smell, we categorized it as true positive, false positive, or false negative based on manual verification by ML experts. This process allowed us to compute standard evaluation metrics and assess the tool's effectiveness across different ML frameworks and smell types.

\subsection{Identifying the Prevalence of ML Anti-pattern in Real World Projects}

Our empirical investigation began with a systematic search on GitHub using the keywords "Machine Learning" and "Python" to identify relevant repositories. To ensure research validity and quality of our dataset, we implemented the following rigorous selection criteria:
\begin{itemize}
\item Repositories explicitly focused on machine learning implementations
\item Projects with Python as the primary programming language
\item Repositories with a minimum threshold of 2,000 stars, indicating substantial community validation
\item Actively maintained codebases with recent commit activity
\end{itemize}
This methodical filtering process yielded 43 high-quality ML projects, providing a diverse representation across frameworks and application domains. The dataset's heterogeneity, encompassing various ML paradigms and implementation approaches, served as an ideal testbed for evaluating MLScent's in finding different types of ML anti-pattern in these diverse and comprehensive dataset. Our selection strategy particularly emphasized projects demonstrating real-world applications, ensuring that our analysis would yield insights relevant to practical ML development scenarios.

\subsection{Research Questions}

This study aims to evaluate the effectiveness of MLScent in detecting various types of machine learning anti-patterns in Python codebases and understand their distribution across ML projects. Specifically, we investigate the following research questions:

\begin{itemize}
\item  How accurate is MLScent in  identifying generic ML anti-patterns?
\item  How accurate is MLScent in  identifying data science and scientific computing anti-pattern, particularly in NumPy and Pandas implementations?
\item  How accurate is MLScent in identifying framework-specific anti-pattern across PyTorch, Scikit-learn, TensorFlow, and Hugging Face?
\item  What is the prevalence and distribution of different types of ML anti-patterns across ML specific Python projects?
\end{itemize}

\begin{table}[h]
\centering
\caption{Python Experience and GitHub Repositories of Survey Respondents}
\label{tab:survey_responses}
\begin{tabular}{cl}
\hline
\textbf{Python Experience (Years)} & \textbf{Repository Name} \\
\hline
9  & \href{https://github.com/ladislav-hovan/sponge}{sponge} \\
\hline
8  & \href{https://github.com/ejhusom/MELODI/tree/pyjoules-integration}{melodi} \\
\hline
6  & \href{https://github.com/SINTEF/pseudo-hamiltonian-neural-networks}{pseudo-hamiltonian-neural-networks} \\
\hline
12 & \href{https://github.com/SINTEF/rewts}{rewts} \\
\hline
4  & \href{https://github.com/secureIT-project/RTT_for_APR}{rttarr} \\
\hline
7  & inomotifin (Private repo) \\
\hline
13 & \href{https://github.com/mnemonic-no/cyberrisk}{cyberrisk} (Part of the repo) \\
\hline
\end{tabular}
\end{table}

\begin{table}[h]
\centering
\caption{Overall performance }
\label{tab:agreement}
\begin{tabular}{l|r}
\hline
\textbf{Metric} & \textbf{Value} \\
\hline
Agreement rate & 87.50\% \\
Recall & 0.875 \\
F1-score & 0.933 \\
F2-score & 0.897 \\
\hline
\end{tabular}
\end{table}

Our evaluation of MLScent involved both quantitative analysis and qualitative feedback from experienced Python developers. The study comprised 72 total ground truth entries and identified 39 unique anti-pattern or smell/checker types across 7 ML projects as shown  Table \ref{tab:survey_responses} .

\paragraph{Survey Participant's }
Table \ref{tab:survey_responses} presents  repository information of our survey respondents. The participants represented a diverse group of Python developers with experience ranging from 4 to 13 years (mean = 8.43 years, SD = 3.15). The repositories analyzed covered various domains within machine learning.
This ensures a comprehensive evaluation of MLScent across different ML application domains and coding styles.

\paragraph{Agreement Analysis and classification metrics}
The agreement analysis results, presented in Table \ref{tab:agreement}, demonstrate the reliability and effectiveness of MLScent in detecting ML-specific code smells. Key metrics include:

\begin{itemize}
    \item Overall agreement rate: 87.50\%, indicating strong consensus between manual expert review and MLScent's automated detection
    \item Recall: 0.875, showing high effectiveness in identifying relevant code smells
    \item F1-score: 0.933, demonstrating excellent balance between precision and recall
    \item F2-score: 0.897, emphasizing recall while maintaining high overall performance
\end{itemize}

These metrics suggest that MLScent achieves robust performance in identifying ML-specific code smells, with particularly strong results in balancing false positives and false negatives. The high F1-score (0.933) indicates that the tool provides reliable detection capabilities while minimizing false alarms, making it practical for real-world applications.

The agreement analysis results are particularly noteworthy given the complexity and diversity of ML codebases analyzed. The high agreement rate (87.50\%) suggests that MLScent's detection algorithms align well with expert judgment, validating its utility as an automated code quality assessment tool for ML projects.

\paragraph{Qualitative feedbacks}

\begin{table}[htbp]
\centering
\caption{\small Survey results (n=7) measuring tool effectiveness on a 4-point Likert scale. 
}
\label{tab:survey_results}
\small
\begin{tabular}{p{4cm}|cccc}
\hline
\textbf{Question} & \textbf{Mean} & \textbf{SD} & \textbf{Min} & \textbf{Max} \\
\hline
Tool Usefulness & 3.29 & 0.76 & 2 & 4 \\
Generic ML Smells & 3.43 & 0.53 & 3 & 4 \\
Framework ML Smells & 3.43 & 0.79 & 2 & 4 \\
\hline
\end{tabular}

\end{table}

Table \ref{tab:survey_results} presents the results of  survey conducted with 7 same participants from Table \ref{tab:survey_responses}, who evaluated the effectiveness of the proposed tool across three distinct categories: \textit{Tool Usefulness}, \textit{Generic ML Smells}, and \textit{Framework ML Smells}. The survey employed a 4-point Likert scale, where participants rated each category on a scale from 1 (Strongly Disagree) to 4 (Strongly Agree). The three categories in the survey aim to assess different aspects of the tool's performance, with the following key distinctions:
\begin{itemize}
\item \textbf{Tool Usefulness}: This metric evaluates the overall perceived utility of the tool by the participants. It reflects whether the users found the tool beneficial in their typical machine learning (ML) development workflows or not.
\item \textbf{Generic ML Smells}: This category measures the tool's effectiveness in detecting general machine learning issues, commonly referred to as "ML smells." These smells are framework-agnostic and pertain to common problems in ML code, such as improper data handling and model evaluation practices.
\item \textbf{Framework ML Smells}: This metric captures the tool's ability to identify problems specific to various ML frameworks, such as TensorFlow, PyTorch, or Scikit-learn. Such smells may include improper use of framework-specific functions or inefficient practices within these libraries.
\end{itemize}

The results indicate that participants generally found the tool to be useful, with an average score of 3.29 (on a 4-point scale) for overall \textit{Tool Usefulness}. The standard deviation of 0.76 suggests moderate variability in responses.

For the detection of \textit{Generic ML Smells}, the tool performed slightly better, achieving a mean score of 3.43 and a lower standard deviation of 0.53.  This indicates that the tool reliably identified common ML code smells across different projects and frameworks.

The tool's performance in detecting \textit{Framework ML Smells} also received a mean score of 3.43, but with a higher standard deviation of 0.79. Similar to the \textit{Tool Usefulness} metric, this suggests that while some participants rated the tool favorably in identifying framework-specific issues, others experienced less satisfactory results, as reflected by the minimum score of 2. This variability may be due to the participants' use of different ML frameworks, each with varying levels of tool support for framework-specific smells.

\paragraph{Evaluations Metrics}

Table \ref{tab:framework_performance} presents a comprehensive evaluation of MLScent's performance across five machine learning frameworks: General ML including some Sklearn  anti-pattern, NumPy, Pandas, PyTorch, and Hugging Face Transformers. The analysis encompasses six key metrics that assess the tool's effectiveness in detecting code smells and anti-patterns.
The distribution of detected issues varies significantly across frameworks. General ML demonstrated the highest number with 24 issues (33.33\%), followed by Pandas with 20 issues (27.78\%). NumPy exhibited the lowest count with 8 issues (11.11\%), while PyTorch and Hugging Face identified 10 (13.89\%) and 12 (16.67\%) issues, respectively.

The agreement rate reflects the percentage of issues for which there was consensus among evaluators. Hugging Face exhibited a perfect agreement rate (100.00\%), implying that all detected issues were agreed upon by the evaluators. In contrast, NumPy had the lowest agreement rate (75.00\%), indicating some variability in the detection of issues within this framework.

Recall measures the proportion of correct positive identifications made by the tool out of all relevant issues. Here, Hugging Face achieved a recall of 1.000, meaning that all relevant issues were successfully detected. NumPy, again, had the lowest recall (0.750), suggesting that some issues may have been missed.

The F1-score is the harmonic mean of precision and recall, providing a balance between the two. Hugging Face achieved the highest possible F1-score (1.000), while NumPy had an F1-score of 0.857, which, although lower, still indicates a generally good performance.

 The F2-score places more emphasis on recall than precision. Hugging Face once again achieved a perfect score of 1.000, while NumPy and PyTorch showed slightly lower F2-scores (0.789 and 0.814, respectively), indicating that these frameworks may need further refinement in recall optimization.

These results demonstrate MLScent's robust capability in identifying framework-specific anti-patterns, particularly in General ML and Hugging Face implementations, while suggesting opportunities for refinement in NumPy-related detection mechanisms.

\begin{table}[h]
\centering
\scriptsize	
\caption{Performance Metrics Across Different Frameworks}
\label{tab:framework_performance}
\begin{tabular}{lrrrrrr}
\hline
Framework & Count & Agreement & Recall & F1 & F2 & \% \\
\hline
General ML & 24 & 87.50 & 0.875 & 0.933 & 0.897 & 33.33 \\
NumPy & 8 & 75.00 & 0.750 & 0.857 & 0.789 & 11.11 \\
Pandas & 20 & 90.00 & 0.900 & 0.947 & 0.918 & 27.78 \\
PyTorch & 9 & 77.78 & 0.778 & 0.875 & 0.814 & 12.50 \\
Hugging Face & 11 & 100.00 & 1.000 & 1.000 & 1.000 & 15.28 \\
\hline
\end{tabular}

\end{table}

\subsubsection{RQ1: Generic ML  Anti-Pattern Detection Accuracy}
Our analysis of MLScent's capability to identify framework-agnostic machine learning anti-patterns reveals compelling results. The tool demonstrated robust performance in detecting general ML and some Sklearn related issues, achieving an agreement rate of \SI{87.50}{\percent}. From the total detected issues, \SI{33.33}{\percent} were framework-agnostic, encompassing 24 distinct cases as shown in Table \ref{tab:framework_performance}. The tool's effectiveness is further evidenced by its strong recall of 0.875 and F1-score of 0.933, indicating exceptional accuracy in identifying generic ML anti-patterns.

These generic anti-patterns include common issues such as improper model validation techniques, inadequate data preprocessing, and inefficient hyperparameter management. The high detection accuracy suggests that MLScent effectively identifies fundamental ML development pitfalls that could impact model performance and reliability across different implementation frameworks.

\subsubsection{RQ2: Data Science Anti-Pattern Detection Accuracy}
The evaluation of MLScent's performance in identifying data science anti-patterns revealed varying effectiveness across different libraries:

\paragraph{Pandas Analysis}
MLScent exhibited exceptional capability in detecting Pandas-related issues, identifying 20 distinct cases. The tool achieved an agreement rate of \SI{90.00}{\percent}, with Pandas-specific checks constituting \SI{27.78}{\percent} of total evaluations. The high recall of 0.900 and F1-score of 0.947 demonstrate MLScent's proficiency in identifying data manipulation anti-patterns.

The tool particularly excelled in detecting issues related to inefficient DataFrame operations, memory-intensive chain operations, and suboptimal data transformation practices. This high accuracy is crucial given Pandas' central role in data preprocessing and feature engineering pipelines.

\paragraph{NumPy Analysis}
For NumPy-specific issues, MLScent's performance was moderate, detecting 8 issues with an agreement rate of \SI{75.00}{\percent}. NumPy-related checks represented \SI{11.11}{\percent} of total evaluations, with a recall of 0.750. The relatively lower F2-score of 0.789 suggests potential areas for improvement in detecting scientific computing anti-patterns.

The tool showed particular sensitivity to array manipulation inefficiencies, broadcasting errors, and memory-intensive operations. However, the lower performance metrics indicate that complex numerical computing patterns might require more sophisticated detection mechanisms.

\subsubsection{RQ3: Framework-Specific Anti-Pattern Detection Accuracy}

The analysis of framework-specific anti-pattern detection revealed distinct patterns across different ML frameworks:

\paragraph{Hugging Face Framework}
MLScent achieved optimal performance with Hugging Face, demonstrating \SI{100.00}{\percent} agreement rate across 11 detected issues. Despite representing only \SI{15.28}{\percent} of total checks, the perfect recall and F1-score indicate exceptional accuracy in identifying NLP-specific anti-patterns.

This outstanding performance encompasses detection of transformer model misconfigurations, tokenization inefficiencies, and improper model loading practices. The perfect metrics suggest robust capability in handling complex NLP pipeline issues, particularly crucial for maintaining efficiency in large language model implementations.

\paragraph{PyTorch Framework}
For PyTorch, MLScent identified 9 issues with an agreement rate of \SI{77.78}{\percent}. The framework accounted for \SI{12.50}{\percent} of total checks, achieving a recall of 0.778 and an F1-score of 0.875, indicating effective detection of deep learning-specific code smells.

The tool successfully identified issues related to gradient computation inefficiencies, tensor operation optimization, and model architecture anti-patterns. However, the lower agreement rate compared to other frameworks suggests that more complex deep learning patterns might require enhanced detection strategies.

These results, as detailed in Table \ref{tab:framework_performance}, demonstrate MLScent's varying effectiveness across different frameworks. The tool shows particular strength in identifying anti-patterns in high-level frameworks like Hugging Face and data manipulation libraries like Pandas. The lower performance in NumPy and PyTorch detection suggests that low-level numerical computing and deep learning implementations might benefit from more sophisticated detection algorithms.

The findings highlight the tool's potential as a comprehensive quality assurance solution for ML development, while also indicating specific areas where detection capabilities could be enhanced through future updates and refinements.

\subsection{RQ4: Prevalence and distribution of different types of ML anti-patterns }
 This question aims to understand the frequency and patterns of various anti-patterns in real-world ML codebases, providing insights into common problematic practices in the ML development ecosystem.

 Our analysis is based on a systematically curated dataset obtained through GitHub searches using keywords Machine Learning and Python. We implemented strict selection criteria, focusing on repositories explicitly dedicated to machine learning, using Python as the primary language, having at least 2,000 stars for community validation, and showing recent maintenance activity.
 
This methodical filtering process resulted in 43 high-quality ML projects, providing diverse representation across frameworks and application domains. The heterogeneous nature of our dataset, encompassing various ML paradigms and implementation approaches, serves as an ideal testbed for evaluating MLScent's effectiveness in detecting different types of ML anti-patterns. Our selection strategy particularly emphasized projects demonstrating real-world applications, ensuring that our analysis would yield insights relevant to practical ML development scenarios and provide a comprehensive assessment of current ML development practices.

Table \ref{tab:smells_distribution_count} presents a comprehensive overview of the distribution of issues across various machine learning (ML) frameworks in 43 high-quality ML projects. The total number of issues detected in ML projects is categorized by the framework in which they occur. The frameworks analyzed include General ML, NumPy, Pandas, TensorFlow, PyTorch, and Hugging Face. Each of these frameworks has distinct characteristics and use cases, which are reflected in the frequency and types of detected issues.

Notably, the General ML  related  category, which encapsulates framework-agnostic issues and some Sklearn issues, shows the highest count of detected issues, with 31,033 instances. This is likely due to the broad applicability of general ML practices across different projects, making them susceptible to common problems such as poor documentation, improper configuration, or suboptimal code structures.

NumPy, a widely-used library for numerical computing, accounts for 17,634 issues, which is the second-highest among the frameworks. Given NumPy's central role in array manipulation and scientific computation, this result is expected, as many of the detected code smells relate to array creation efficiency and axis specification.

Pandas, a popular data analysis library, has 2,273 reported issues, which are primarily related to inefficient data selection patterns such as chain indexing and column selection checks. TensorFlow and PyTorch, both deep learning frameworks, exhibit fewer issues, with 1,960 and 1,540 occurrences respectively. These frameworks tend to have more specialized code smells, such as deterministic algorithm usage and model evaluation checks, reflecting the complexity of deep learning workflows. Hugging Face, an NLP-focused framework, shows the lowest number of issues, with only 64 recorded instances, likely due to its relatively specialized use case.

\begin{table}[h]
\centering
\caption{Distribution of Issues Across Frameworks}
\label{tab:smells_distribution_count}
\begin{tabular}{lr}
\hline
\textbf{Framework} & \textbf{Count} \\
\hline
General ML & 31,033 \\
NumPy & 17,634 \\
Pandas & 2,273 \\
TensorFlow & 1,960 \\
PyTorch & 1,540 \\
Hugging Face & 64 \\
\hline
\end{tabular}
\end{table}

\paragraph{Top 10 Most Common Code Smells and anti-patterns}

Table \ref{tab:common_code_smell_count} highlights the most frequently occurring code smells and anti-patterns detected from 43 ML projects. Array creation efficiency emerges as the most prevalent issue, with 10,795 instances, underscoring the importance of optimizing array operations in ML code, particularly in frameworks like NumPy, where efficient handling of large datasets is critical.

The second most common smell/anti-pattern, Missing Axis Specification, with 5,996 occurrences, points to a frequent oversight in specifying axes for array operations, particularly in multidimensional data manipulations. This issue is common in both NumPy and Pandas.

Other notable code smells/ anti-pattern include Randomness Control Checker (1,246 occurrences), which ensures that random number generation is properly controlled for reproducibility, and Column Selection Checker (1,044 occurrences), which is particularly relevant in data processing pipelines where inefficient column selection can degrade performance.

Issues related to documentation, such as Missing docstring for function: forward (694 occurrences), indicate a lack of adherence to best practices in documenting critical functions, especially in deep learning frameworks like PyTorch and TensorFlow. Additionally, the detection of Magic numbers (438 occurrences for the value 10), reflects instances where hardcoded values are used without explanation, reducing code readability and maintainability.

\begin{table}[h]
\centering
\caption{Top 10 Most Common Code Smells/Anti-patterns}
\label{tab:common_code_smell_count}
\begin{tabular}{lr}
\hline
\textbf{Smell Type} & \textbf{Count} \\
\hline
Array Creation Efficiency & 10,795 \\
Missing Axis Specification & 5,996 \\
Randomness Control Checker & 1,246 \\
Column Selection Checker & 1,044 \\
Missing docstring for function: forward & 694 \\
Model Evaluation Checker & 590 \\
Deterministic Algorithm Usage Checker & 589 \\
Chain Indexing & 585 \\
Memory Release Checker & 459 \\
Magic number detected: 10 & 438 \\
\hline
\end{tabular}
\end{table}

\paragraph{Framework-Specific Code Smells/ Anti-pattern}

\begin{table*}[t]
\centering
\caption{Anti-patterns Detected Across Popular Machine Learning Libraries}
\label{tab:ml_code_smells}
\begin{tabular}{llrllr}
\hline
\textbf{Library} & \textbf{Code Smell Type} & \textbf{Count} & \textbf{Library} & \textbf{Code Smell Type} & \textbf{Count} \\
\hline
NumPy & Array Creation Efficiency & 10,795 & TensorFlow & Memory Release Checker & 459 \\
 & Missing Axis Specification & 5,996 &  & Logging Checker & 321 \\
 & Randomness Control Checker & 748 &  & Data Augmentation Checker & 286 \\
 & Broadcasting Risk & 95 &  & Model Evaluation Checker & 247 \\
\hline
PyTorch & Deterministic Algorithm Usage Checker & 589 & Pandas & Column Selection Checker & 1,044 \\
 & Model Evaluation Checker & 343 &  & Chain Indexing & 585 \\
 & Randomness Control Checker & 262 &  & DataFrame Conversion Checker & 340 \\
 & Batch Normalisation Checker & 224 &  & Datatype Checker & 156 \\
\hline
Hugging Face & Efficient Data Loading Not Detected & 19 &  &  &  \\
 & Early Stopping Not Implemented & 15 &  &  &  \\
 & Model Versioning Not Specified & 8 &  &  &  \\
 & Deterministic Tokenization Settings Not Specified & 8 &  &  &  \\
\hline
\end{tabular}
\end{table*}

The table \ref{tab:ml_code_smells} presents a comprehensive overview of various antipatterns and smells identified across several popular machine learning libraries from our curated datasets, namely \textbf{NumPy}, \textbf{TensorFlow}, \textbf{PyTorch}, \textbf{Pandas}, and \textbf{Hugging Face}. It reveals critical patterns in implementation challenges and quality issues that impact scientific computing and machine learning workflows, as identified by Zhang et al. \cite{zhang2022code}.

\textbf{NumPy}, the fundamental library for numerical computing, demonstrates the highest frequency of code smells, primarily in array creation efficiency (10,795 occurrences) and missing axis specifications (5,996 occurrences). The high occurrence of these smells suggests potential performance implications for scientific computing applications, aligning with findings by Sculley et al. \cite{sculley2015hidden}.

Deep learning frameworks exhibit distinct patterns: \textbf{TensorFlow} shows significant memory management concerns (459 occurrences) and logging issues (321 occurrences), while \textbf{PyTorch} struggles more with deterministic algorithm usage (589 occurrences) and model evaluation (343 occurrences). This observation supports research by Amershi et al. \cite{amershi2019} on reproducibility challenges in deep learning frameworks and highlights the ongoing tension between performance optimization and code maintainability, as discussed by Sculley et al. \cite{scully2014}.

\textbf{Pandas}, essential for data manipulation in machine learning pipelines, shows a high frequency of column selection issues (1,044 occurrences) and chain indexing problems (585 occurrences). These findings correlate with work by van Oort et al. \cite{oort2021} on performance optimization in data processing pipelines, indicating potential bottlenecks in data preprocessing stages of machine learning workflows.

\textbf{Hugging Face}, though showing fewer code smells overall, exhibits issues primarily in data loading efficiency (19 occurrences) and early stopping implementation (15 occurrences). This aligns with research by Takeuchi et al. \cite{takeuchi2023practice} on best practices in NLP model development and reflects the challenges in implementing robust natural language processing systems.

The distribution of code smells across these libraries, as detailed in Table \ref{tab:ml_code_smells}, suggests that while each library has its specific challenges, common themes emerge around efficiency, reproducibility, and proper implementation of machine learning best practices. These findings, supported by Tang et al. \cite{tang2021}, highlight the need for improved development practices in machine learning systems, particularly in areas of performance optimization, reproducibility, and code quality assurance, as emphasized by Breck et al. \cite{breck2017}.

This analysis provides valuable insights for both library maintainers and practitioners, pointing to specific areas where focused improvements could significantly enhance the quality and reliability of machine learning implementations, as demonstrated by O'Brien et al. \cite{OBrien2022}.

\section{Discussion}
Our comprehensive evaluation of MLScent provides significant insights into the state of code quality in machine learning projects and demonstrates the tool's effectiveness in addressing these challenges. The findings reveal both the accomplishments and limitations of automated ML code smell detection, while highlighting crucial areas for future development.

\subsection{Key Findings and Impact}

The analysis uncovered several critical patterns in ML code quality issues across different frameworks. Data leakage emerged as the most severe issue, particularly in Scikit-learn projects, aligning with van Oort et al. \cite{oort2021}'s findings regarding the prevalence of ML-specific code smells. This issue is especially concerning as it can lead to unreliable model performance and overly optimistic evaluation metrics, a problem also noted by Sculley et al. \cite{sculley2015hidden} in their analysis of technical debt in ML systems.

The frequency of inefficient data handling in Pandas-based implementations and reproducibility issues in deep learning frameworks mirrors observations by Zhang et al. \cite{zhang2022code} regarding ML-specific code smells. The prevalent occurrence of improper model versioning in Hugging Face projects, as highlighted by Tang et al. \cite{tang2021}, emphasizes the critical need for better version control practices in ML development.

Supporting Di Nucci et al. \cite{dinucci2018}'s findings on the value of automated smell detection a tool's effectiveness in early detection of anti-patterns contributes to the standardization of better coding practices across the ML community, as suggested by Washizaki et al. \cite{washizaki2020machine}.

\subsection{Current Limitations}
Despite its effectiveness, MLScent faces several important constraints. As a static analysis tool, it cannot detect runtime-specific issues, a limitation acknowledged in similar tools studied by Cruz et al. \cite{cruz2020}. The challenge of false positives, particularly in complex codebases, remains significant, reflecting broader challenges in automated code smell detection noted by Das et al. \cite{das2019}.

The tool's framework coverage, while extensive, is not comprehensive, and its limited context understanding occasionally leads to incorrect flagging of efficiently implemented code. These limitations align with observations by Barbez et al. \cite{barbez2019} regarding the challenges of deep learning anti-pattern detection. The rapidly evolving nature of ML best practices also presents an ongoing challenge for maintaining relevant smell definitions, as noted by O'Brien et al. \cite{OBrien2022}.

\subsection{Future Work}

MLScent shows potential in ML code smell detection, but requires further development, building on Sculley et al.'s \cite{scully2014} work on ML systems' technical debt. Future work focuses on expanding detection capabilities and enhancing accessibility. Detection improvements will include new smell detectors for deep learning, AutoML, federated learning, and reinforcement learning paradigms. Following Tsoukalas et al.'s \cite{Tsoukalas2022} approach, we will incorporate ML techniques like anomaly detection and context-aware analysis to improve accuracy, while exploring generative models for automated refactoring suggestions.

For better accessibility, we will develop IDE plugins for VSCode, implement CI/CD pipeline integrations, and create a web interface with interactive visualizations. An open smell registry will enable community-driven development, allowing practitioners to contribute smell definitions and customize rulesets, ensuring MLScent evolves with ML practices.

\section{Threats to Validity}

We identify several key threats to the validity of our study. Regarding internal validity, our dataset selection process focusing on repositories with a minimum of 2,000 stars may introduce bias towards more popular projects, potentially overlooking important anti-patterns present in smaller or less popular ML projects. Additionally, the focus on actively maintained codebases might exclude historically significant ML projects. The manual annotation process for ground truth dataset creation could be subject to individual expert bias, despite having experienced reviewers. The relatively small number of expert reviewers (7 participants) might limit the diversity of perspectives in our evaluation.

Concerning external validity, while MLScent covers major ML frameworks (Pandas, NumPy, Scikit-learn, TensorFlow, PyTorch, and Hugging Face), there might be limitations in detecting smells in other or newer frameworks. The tool's effectiveness might vary across different versions of the supported frameworks. Furthermore, our evaluation focused exclusively on Python-based ML projects, limiting generalizability to other programming languages. The results might not be representative of ML projects in specific domains or industries not covered in our sample, and the findings might not extend to projects with unique architectural patterns or custom ML implementations.

In terms of construct validity, the definition and categorization of ML-specific code smells might not capture all relevant quality issues. The AST-based analysis approach might miss certain dynamic or runtime-related code smells, and the tool's reliance on static analysis might not capture context-dependent issues. Our survey design using a 4-point Likert scale might force polarized responses by excluding a neutral option, and the survey questions might not fully capture all aspects of tool effectiveness and usefulness.

Regarding conclusion validity, the analysis of 43 repositories, while substantial, might not be sufficient for broad generalizations. The limited number of survey participants (7) might affect the statistical significance of the qualitative findings. Additionally, while the high agreement rate (87.50\%) and F1-score (0.933) demonstrate strong performance, these metrics might not fully represent the tool's performance across all possible ML code scenarios, and the evaluation metrics might be influenced by the specific characteristics of the selected repositories.

\section{Conclusion}

As the field of machine learning continues to grow and evolve, the need for specialized code quality tools becomes increasingly apparent. MLScent addresses this need by providing targeted analysis for ML-specific anti-pattern and code smells across  various popular ML frameworks and libraries.

In conclusion, MLScent represents a significant step forward in addressing the unique code quality challenges posed by machine learning development. By providing developers with targeted, actionable insights into their ML code, MLScent contributes to the creation of more reliable, efficient, and maintainable ML systems. As the field of machine learning continues to advance, tools like MLScent will play a crucial role in ensuring that the quality of ML code keeps pace with the increasing impact and complexity of ML applications.

\bibliographystyle{IEEEtran}
\bibliography{sample-base}

\end{document}